\documentclass[aps,prd,twocolumn,10pt,showpacs,superscriptaddress,preprintnumbers,nofootinbib,amsmath,amssymb,floatfix]{revtex4-2}

\usepackage[colorlinks=true,urlcolor=blue,linkcolor=blue,citecolor=blue]{hyperref}
\usepackage[normalem]{ulem}
\usepackage[capitalize]{cleveref}
\usepackage{mathtools}
\usepackage{microtype}
\usepackage{xcolor}
\usepackage{graphicx}
\usepackage{dcolumn}
\usepackage{bm}
\usepackage{multirow}

\newcommand{\Beq}{\begin{equation}\begin{aligned}}
\newcommand{\Eeq}{\end{aligned}\end{equation}}
\newcommand{\mpl}{m_\mathrm{Pl}}
\newcommand{\mA}{m_{\gamma'}}

\begin{document}

\title{Dark photon dark matter from an oscillating dilaton}


\author{Peter Adshead}
\email{adshead@illinois.edu}
\affiliation{Illinois Center for Advanced Studies of the Universe \& Department of Physics, University of Illinois at Urbana-Champaign, Urbana, IL 61801, USA,}
\affiliation{Center for Particle Cosmology, Department of Physics and Astronomy, University of Pennsylvania,
209 South 33rd St, Philadelphia, PA 19104}
\author{Kaloian D. Lozanov}
\email{kaloian.lozanov@ipmu.jp}
\affiliation{Illinois Center for Advanced Studies of the Universe \& Department of Physics, University of Illinois at Urbana-Champaign, Urbana, IL 61801, USA,}
\affiliation{Kavli IPMU (WPI), UTIAS, The University of Tokyo, 5-1-5 Kashiwanoha, Kashiwa, Chiba 277-8583, Japan}
\author{Zachary J. Weiner}
\email{zweiner@uw.edu}
\affiliation{Department of Physics, University of Washington, Seattle, WA 98195, USA.}


\date{\today}

\begin{abstract}
We present a mechanism for generating ultralight dark photon dark matter in the early Universe via a
dilatonlike scalar field coupled to the dark photon's kinetic term.
Energy is initially stored in the condensate of the dilaton, which resonantly produces dark photons
when it begins oscillating in the early Universe.
While similar scenarios with axion--dark-photon couplings require large coupling coefficients to
fully populate the dark photon, the dilatonic coupling features a unique regime:
When the dark photon's mass is half that of the dilaton, dark photons are copiously produced even
when the dilaton undergoes small-amplitude oscillations.
Scenarios consistent with the cosmic microwave background allow for ultralight vector dark matter with mass as light as $10^{-20}$ eV.
\end{abstract}


\maketitle

\section{Introduction}

Ultralight, massive dark photons (i.e., spin-1 vector bosons) are a curious candidate for the dark matter (DM) in our Universe. Like scalar fuzzy DM~\cite{Hu:2000ke,Schive:2014dra,Hui:2016ltb,Hui:2021tkt}, ultralight dark photons exhibit wavelike properties on macroscopic scales, $\lambda\lesssim 10$~pc, for masses $\mA\gtrsim 10^{-21}\,\mathrm{eV}$~\cite{Adshead:2021kvl,Salehian:2021khb}. Halos supported by dark photons can therefore feature vector solitonic cores~\cite{Adshead:2021kvl} reminiscent of those in scalar fuzzy DM models~\cite{Hu:2000ke}. However, these vector solitons are distinguished due to their intrinsic spins~\cite{Jain:2021pnk,Amin:2022pzv}. The vectorlike nature of the condensate also allows for distinctive higher energy solitonic configurations similar to Proca stars with radially directed vector fields~\cite{Adshead:2021kvl}. Furthermore, stable starlike solutions are possible when self-interactions of the dark photons are included~\cite{Zhang:2021xxa,Jain:2022kwq}. Vector dark matter's predictions for small-scale structure also differ considerably from its scalar cousin due to the nature of its production mechanisms, which typically result in a highly peaked power spectrum on small scales~\cite{Graham:2015rva, Deskins:2013dwa, Adshead:2015pva, Adshead:2018doq, Agrawal:2018vin}. Such a peaked spectrum leads to rich small-scale structure, with significant energy density stored in boson stars~\cite{Gorghetto:2022sue, Amin:2022pzv}.

Recent studies on dynamical heating of ultrafaint dwarf galaxies via coherent fluctuations in fuzzy dark matter put pressure on the very low mass end of the mass spectrum, requiring $m_\mathrm{fdm} > 10^{-19}$ eV~\cite{Dalal:2022rmp}.
Slightly weaker limits are expected to apply to the vector scenario due to the reduced interference
of the multiple polarization states~\cite{Amin:2022pzv}.
Reference~\cite{Amin:2022nlh} argues that a generic bound $m_\mathrm{fdm} > 10^{-18} \, \mathrm{eV}$
applies when dark matter is produced by a causal process after inflation.
While scalar~\cite{Arvanitaki:2009fg, Arvanitaki:2010sy, Arvanitaki:2016qwi, Baryakhtar:2020gao} and vector~\cite{Baryakhtar:2017ngi, Cardoso:2018tly} masses in the range $10^{-13}-10^{-11}$ eV are constrained by solar-mass black hole superradiance bounds (and lighter masses could be probed by supermassive black holes~\cite{Baryakhtar:2020gao}), these can be evaded by self-interactions of the vector field~\cite{Agrawal:2018vin}, leaving a wide range of masses viable.

Production of dark photon DM in the ultralight particle mass range is a longstanding problem. Early models that produced dark photons from a misalignment mechanism analogous to that of scalar dark matter production~\cite{Nelson:2011sf} were later shown to require nonminimal couplings to the Ricci scalar~\cite{Arias:2012az, Alonso-Alvarez:2019ixv, Elahi:2022hgj} that can lead to violations of unitarity at relatively low energy scales in longitudinal graviton-photon scattering~\cite{Agrawal:2018vin}. Massive dark photons minimally coupled to Einstein gravity (but otherwise decoupled from other matter fields) are produced during slow-roll inflation, but their abundance matches the one of DM only if $\mA\gtrsim 10^{-5}\,\mathrm{eV}$~\cite{Graham:2015rva,Ema:2019yrd,Kolb:2020fwh,Ahmed:2020fhc}. Finally, an oscillating Higgs~\cite{Dror:2018pdh} or an oscillating, misaligned axion~\cite{Agrawal:2018vin,Co:2018lka} allow for the resonant production of ultralight dark photons with the correct DM abundance. While these resonant vector DM production models are based on well-motivated theories, they pose questions about naturalness and choices of couplings. In particular, the dimensionless coupling constants should respect a hierarchy in the case of the oscillating Higgs~\cite{Dror:2018pdh}, and in the axion models, large couplings between the axion and the Chern-Simons term $F \tilde{F}$ of the vector field are exacerbated by the small gauge couplings that are required~\cite{Agrawal:2018vin,Co:2018lka}.

In this work, we identify a new variant of the resonant mechanism for generating dark photon DM that does not require unnaturally large couplings. In addition to the dark photon, we consider a scalar (or ``dilaton'') $\phi$ kinetically coupled to the dark photon via an interaction $\mathcal{L}\supset W(\phi) F_{\mu \nu} F^{\mu \nu} / 4$.
This form of interaction was studied in Refs.~\cite{Nakai:2020cfw,Nakayama:2020rka}, where $\phi$ played the role of the inflaton. During slow-roll inflation, ultralight dark photons can be produced with the right DM abundance provided that the effective dependence of the coupling on the Friedmann-Lema\^itre-Robertson-Walker (FLRW) scale factor is $W(\phi(a))\propto a^{-n}$ for $n$ close to $4$.
Reference~\cite{Nakai:2022dni} considered a spectator $\phi$ whose evolution during inflation produces
large-wavelength dark photons.
In this work, we instead consider $\phi$ to be a light spectator field during inflation whose
postinflationary oscillations (rather than its inflationary dynamics) give rise to the resonant
production of vector DM.

When the dilaton oscillates with large amplitude [$\phi/M \gg 1$, where $M$ is the mass scale associated with the coupling function $W(\phi)$], dark
photons are efficiently produced via broad resonance, closely resembling models that feature a
coupling to an axion~\cite{Agrawal:2018vin,Co:2018lka}.
Typically, such parametric resonances become inefficient for small oscillation amplitudes
$\phi / M < 1$, below which
\begin{align}
    W(\phi)
    &\approx 1
        + \phi / M
        + \mathcal{O}[(\phi/M)^{2}].
    \label{eqn:W-asymptotic}
\end{align}
As such, large couplings are required to completely deplete the dilaton (or axion) condensate into
the vector.
However, the dilatonic coupling exhibits a unique regime. When the vector mass is half the dilaton
mass, $\mA = m_\phi / 2$, a low-momentum instability becomes efficient at late times,
i.e., under \textit{small} amplitude oscillations $\phi < M$.
We show this explicitly for the benchmark model $W(\phi) = e^{\phi/M}$, but our main results depend
only on the small-amplitude behavior being of the form \cref{eqn:W-asymptotic}.
With $M \approx 10^{17} \, \mathrm{GeV}$, the model successfully realizes ultralight vector DM
scenarios ($\mA \sim 10^{-21} \, \mathrm{eV}$) consistent with cosmic microwave background (CMB)
bounds on isocurvature perturbations and the scale of inflation.
Unlike existing resonance mechanisms~\cite{Agrawal:2018vin,Co:2018lka}, this mechanism therefore
does not require a large coupling between the dark photon and the scalar, instead requiring a
specific tuning between the masses of the scalar and dark photon.

In the remainder of this paper, we detail the governing equations for a massive vector coupled to a
dilaton (\cref{sec:setup}), study the resonant production of dark photons due to an oscillating
dilaton (\cref{sec:StabAn}), and present the resulting relic abundance of dark photon dark matter and viable parameter space given various cosmological constraints (\cref{sec:DMAbun}).
We conclude in \cref{sec:Concl}.
Throughout, we work with an FLRW metric of the form
\begin{align}\label{eqn:cosmic-time-flrw-metric}
    \mathrm{d} s^2
    &= - \mathrm{d} t^2 + a(t)^2 \delta_{i j} \mathrm{d} x^i \mathrm{d} x^j,
\end{align}
with $a(t)$ the scale factor.
We use natural units in which $\hbar = c = 1$ and the reduced Planck
mass $\mpl = 1/\sqrt{8\pi G}$.
Greek spacetime indices are contracted via the Einstein summation convention, while repeated Latin
spatial indices are contracted with the Kronecker delta function (regardless of their placement).
Dots denote derivatives with respect to cosmic time $t$, and the Hubble rate is $H \equiv \dot{a} / a$.

\section{Model and dynamics}
\label{sec:setup}

We consider a dilatonlike scalar $\phi$ coupled to a vector $A_\mu$ via the action
\begin{align}
\label{eq:L}
\begin{split}
    S
    &= \int \mathrm{d}^4 x \sqrt{-g}
        \Bigg[
            \frac{\mpl^2}{2} R
            - \frac{1}{2} \partial_\mu \phi \partial^\mu \phi
            - \frac{1}{2} m_\phi^2 \phi^2
    \\ &\hphantom{ {}={} \int \mathrm{d}^4 x \sqrt{-g} \Bigg[ }
            - \frac{W(\phi)}{4} F_{\mu\nu} F^{\mu\nu}
            - \frac{1}{2} \mA^2 A_\mu A^\mu
        \Bigg],
\end{split}
\end{align}
where $F_{\mu \nu} = \partial_\mu A_\nu - \partial_\nu A_\mu$ and $R$ is the Ricci scalar.
The modulation of the dark photon kinetic term by $W(\phi)$ can be interpreted as a
$\phi$ dependence of the dark $U(1)$ coupling strength.
The dark photon mass term could arise through the Stueckelberg or Higgs mechanisms.
In general, $\mA$ could be a function of $\phi$, but its $\phi$ dependence need not coincide with
the one of the dark photon gauge coupling.
For simplicity, throughout this work we treat $\mA$ as a constant Stueckelberg/Proca mass term.

The Euler-Lagrange equations for \cref{eq:L} are
\begin{subequations}
\begin{align}
    0
    &= - \nabla_\mu \nabla^\mu \phi
        + m_\phi^2 \phi
        + \frac{W'(\phi)}{4} F_{\mu\nu} F^{\mu\nu}
    \label{eq:dilaton-eom}
    \\
    0
    &= - \nabla_\mu \left[ W(\phi) F^{\mu\nu} \right]
        + \mA^2 A^\nu.
    \label{eq:vector-eom}
\end{align}
\end{subequations}
\Cref{eq:vector-eom} implies that
\begin{align}\label{eqn:lorenz}
    \nabla_\mu A^\mu
    &= 0,
\end{align}
which coincides with the Lorenz gauge choice (for gauge theories) but is instead here a
constraint imposed by consistency of the equations of motion.
In FLRW spacetime, \cref{eqn:cosmic-time-flrw-metric,eq:dilaton-eom,eq:vector-eom} reduce
to
\begin{subequations}
\begin{align}
    0
    &= \ddot{\phi}
        + 3 H \dot{\phi}
        - \frac{1}{a^2} \partial_i \partial_i \phi
        + m_\phi^2 \phi
        + \frac{W'(\phi)}{4} F_{\mu\nu} F^{\mu\nu}
    \label{eqn:dilaton-eom}
    \\
\begin{split}
    0
    &= \ddot{A}_0
        + 3 H \dot{A}_0 + 3 \dot{H} A_0
        - \frac{1}{a^2} \partial_j \partial_j A_0
        + \frac{2 H}{a^2} \partial_i A_i
    \\ &\hphantom{{}={}}
        + \frac{\mA^2}{W(\phi)} A_0
        - \frac{1}{a^2} \frac{\partial_i W(\phi)}{W(\phi)}
        \left(
            \partial_i A_0 - \dot{A}_i
        \right)
    \label{eqn:A0-eom}
\end{split}
    \\
\begin{split}
    0
    &= \ddot{A}_i + H \dot{A}_i
        - \frac{1}{a^2} \partial_j \partial_j A_i
        + 2 H \partial_i A_0
    \\ &\hphantom{{}={}}
        + \frac{\mA^2}{W(\phi)} A_i
        - \frac{\partial^\mu W(\phi)}{W(\phi)} F_{\mu i}.
    \label{eqn:Ai-eom}
\end{split}
\end{align}
\end{subequations}
We expand the vector in Fourier modes as
\begin{align}
    A_0(t, \mathbf{x})
    &= \int \frac{\mathrm{d}^3 k}{(2 \pi)^3} \,
        e^{i\mathbf{k}\cdot\mathbf{x}} A_0(t, \mathbf{k}) \\
    A_j(t, \mathbf{x})
    &= \sum_{\lambda \in \{\pm, \parallel\}}
        \int \frac{\mathrm{d}^3 k}{(2 \pi)^3}
        e^{i\mathbf{k}\cdot\mathbf{x}}
        A_\lambda(t, \mathbf{k}) \epsilon_{j}^\lambda(\mathbf{k}),
\end{align}
where the polarization vectors satisfy
\begin{equation}
\begin{alignedat}{2}
    \epsilon_{m}^\lambda(\mathbf{k})^\ast
    \epsilon_{m}^{\lambda'}(\mathbf{k})
    &= \delta^{\lambda\lambda'}
    \quad &
    i k_m \epsilon_{m}^{\pm}(\mathbf{k})
    &= 0
    \\
    \epsilon_{m}^\lambda(\mathbf{k})
    &= \epsilon_{m}^\lambda(-\mathbf{k})^\ast
    \quad &
    i k_m \epsilon_{m}^\parallel(\mathbf{k})
    &= k
    \\
    i \varepsilon_{lmn} k_m \epsilon_{n}^{\pm}(\mathbf{k})
    &= \pm \epsilon_{l}^{\pm}(\mathbf{k})
    \quad &
    i \varepsilon_{lmn} k_m \epsilon_{n}^\parallel(\mathbf{k})
    &= 0,
\end{alignedat}
\end{equation}
and we likewise expand the dilaton into
\begin{align}
    \phi(t, \mathbf{x})
    &= \bar{\phi}(t)
        + \int \frac{\mathrm{d}^3 k}{(2 \pi)^3}
        e^{i\mathbf{k}\cdot\mathbf{x}} \delta \phi(t, \mathbf{k}).
\end{align}
To linear order in spatial fluctuations, \cref{eqn:Ai-eom} then decomposes into
\begin{subequations}
\begin{align}
    0
    &= \ddot{A}_\pm
        + \left(
            H
            + \frac{\dot{\bar{W}}}{\bar{W}}
        \right)
        \dot{A}_\pm
        + \left(
            \frac{k^2}{a^2}
            + \frac{\mA^2}{\bar{W}}
        \right) A_\pm
    \label{eqn:Ai-eom-linearized-transverse}
    \\
\begin{split}
    0
    &= \ddot{A}_\parallel
        + \left(
            H
            + \frac{\dot{\bar{W}}}{\bar{W}}
        \right)
        \dot{A}_\parallel
        + \left(
            \frac{k^2}{a^2}
            + \frac{\mA^2}{\bar{W}}
        \right)
        A_\parallel
    \\ &\hphantom{{}={}}
        + \left(
            \frac{\dot{\bar{W}}}{\bar{W}}
            - 2 H
        \right) k A_0,
    \label{eqn:Ai-eom-linearized-longitudinal}
\end{split}
\end{align}
\end{subequations}
using the shorthand $\bar{W} = W(\bar{\phi})$.
The Lorenz constraint \cref{eqn:lorenz} sets
\begin{align}
    k A_\parallel / a^2
    &= \dot{A}_0 + 3 H A_0,
    \label{eqn:lorenz-cosmic-fourier}
\end{align}
which combines with \cref{eqn:A0-eom} to give
\begin{align}
    0
    &= \ddot{A}_0
        + 5 H \dot{A}_0
        + \left(
            \frac{k^2}{a^2}
            + \frac{\mA^2}{\bar{W}}
            + 3 \dot{H} + 6 H^2
        \right)
        A_0.
    \label{eqn:A0-eom-linearized-longitudinal}
\end{align}
Note that the third-order differential equation yielded by substituting \cref{eqn:lorenz-cosmic-fourier} for $A_\parallel$ in \cref{eqn:Ai-eom-linearized-longitudinal}
is proportional to the time derivative of \cref{eqn:A0-eom-linearized-longitudinal}
plus $3 H + \dot{\bar{W}} / \bar{W}$ times \cref{eqn:A0-eom-linearized-longitudinal}; i.e.,
the system is self-consistent.

The dynamics of the transverse modes and $A_0$ are most conveniently studied in terms of the
rescaled fields
\begin{subequations}
\begin{align}
    \mathcal{A}_\pm
    &\equiv \sqrt{a \bar{W}} A_\pm
    \\
    \mathcal{A}_0
    &\equiv a^{5/2} A_0,
\end{align}
\end{subequations}
for which \cref{eqn:Ai-eom-linearized-transverse,eqn:A0-eom-linearized-longitudinal}
respectively reduce to
\begin{subequations}
\begin{align}
    \ddot{\mathcal{A}}_\pm
    &= - \left[
            \frac{k^2}{a^2}
            + \frac{\mA^2}{\bar{W}}
            - \frac{\partial_t^2 \sqrt{\bar{W}}}{\sqrt{\bar{W}}}
            - \frac{H \dot{\bar{W}}}{2 \bar{W}}
            - \frac{\dot{H}}{2}
            - \frac{H^2}{4}
        \right]
        \mathcal{A}_\pm
    \label{eqn:Apm-eom-rescaled}
    \\
    \ddot{\mathcal{A}}_0
    &= - \left(
            \frac{k^2}{a^2}
            + \frac{\mA^2}{\bar{W}}
            + \frac{\dot{H}}{2}
            - \frac{H^2}{4}
        \right)
        \mathcal{A}_0.
    \label{eqn:A0-eom-rescaled}
\end{align}
\end{subequations}
In general, $W(\phi) \to 1$ at late times (as $\phi$ decays with expansion), and $\mathcal{A}_{\pm}$
and $\sqrt{a} A_{\pm}$ then coincide, but an instability analysis is more straightforward in terms
of the former.
In addition, \cref{eqn:lorenz-cosmic-fourier} sets
\begin{align}
    A_\parallel
    &= \frac{1}{\sqrt{a} k}
        \left(
            \dot{\mathcal{A}_0}
            + \frac{H}{2} \mathcal{A}_0
        \right);
\end{align}
i.e., one can study the (relatively simpler) dynamics of \cref{eqn:A0-eom-rescaled} in place of the
longitudinal mode.

\section{Stability analysis}
\label{sec:StabAn}

To characterize the (in)stability of vector modes in an oscillating dilaton background, we first
consider the limiting case of Minkowski spacetime.
We then apply these results to FLRW spacetime, relying on the fact that (for the parameter space of
interest) the relevant dynamics occur on timescales much shorter than the instantaneous Hubble rate.

\subsection{Small-amplitude broad resonance in Minkowski spacetime}\label{sec:small-amplitude}

In Minkowski (i.e., nonexpanding) spacetime, the homogeneous component of the dilaton evolves
according to
\begin{align}
    \ddot{\bar{\phi}}
    + m_\phi^2 \bar{\phi}
    &= 0,
\end{align}
solved by $\bar{\phi}(t) = \phi_0 \cos(m_\phi t)$ under the initial condition
$\bar{\phi}(t) \to \phi_0$ and $\dot{\bar{\phi}}(t) \to 0$ for $t \ll 1 / m_\phi$.
For concreteness, we set
\begin{align}
    W(\phi)
    &= e^{\phi/M},
\end{align}
for which \cref{eqn:A0-eom-rescaled,eqn:Apm-eom-rescaled} (after substituting $a = 1$ and $H = 0$)
are
\begin{subequations}
\begin{align}
    \ddot{\mathcal{A}}_\pm
    &= - \left[
            k^2
            + \frac{\mA^2}{e^{\bar{\phi} / M}}
            - \frac{\ddot{\bar{\phi}}}{2 M}
            - \left( \frac{\dot{\bar{\phi}}}{2 M} \right)^2
        \right]
        \mathcal{A}_\pm
    \label{eqn:Apm-eom-rescaled-minkowski}
    \\
    \ddot{\mathcal{A}}_0
    &= - \left(
            k^2
            + \frac{\mA^2}{e^{\bar{\phi} / M}}
        \right)
        \mathcal{A}_0.
    \label{eqn:A0-eom-rescaled-minkowski}
\end{align}
\end{subequations}
Because $\bar{\phi}(t)$ is periodic,
\cref{eqn:A0-eom-rescaled-minkowski,eqn:Apm-eom-rescaled-minkowski} describe harmonic oscillators
with periodic frequencies.
By the Floquet theorem~\cite{magnus2004hill,Amin:2014eta}, their solutions are of the form
$\mathcal{P}_+(t)e^{\mu t}+\mathcal{P}_-(t)e^{-\mu t}$, where $\mu$ is the Floquet exponent and
$\mathcal{P}_\pm(t)=\mathcal{P}_\pm(t+T)$, with $T$ being the period of the background.
Modes for which $\Re({\mu}) \neq 0$ are unstable and grow exponentially with time.
We next study the structure of parametric instabilities in these equations.

\begin{figure*}[ht!]
    \centering
    \includegraphics[width=\textwidth]{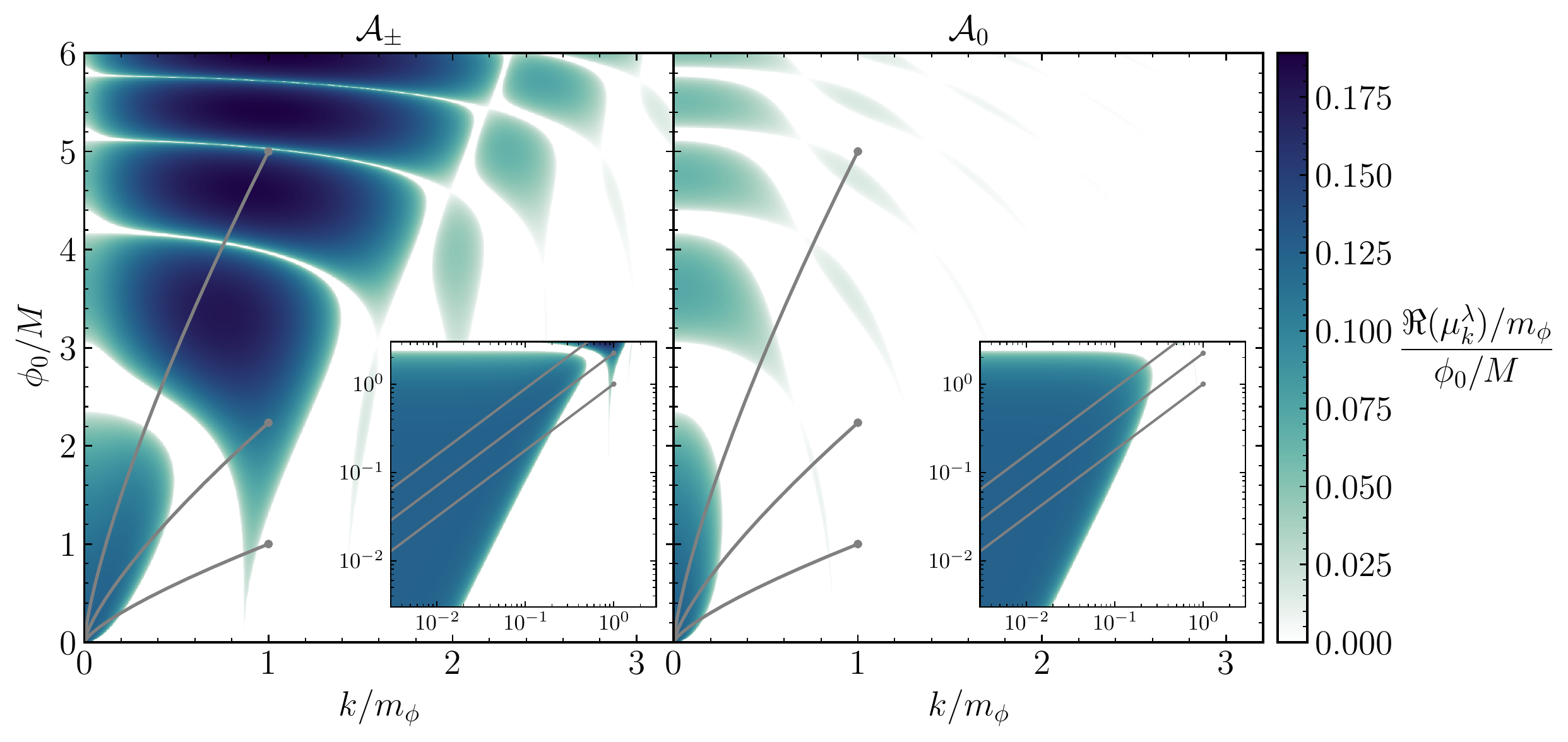}
    \caption{
        Real part of the Floquet exponent $\mu_k$ for the solutions of the equations of motion
        describing the transverse [\cref{eqn:Apm-eom-rescaled-minkowski}] and longitudinal
        [\cref{eqn:A0-eom-rescaled-minkowski}] sectors.
        Both panels fix $m_\phi = 2 \mA$, for which a region with $\Re(\mu_k) > 0$ extends to
        arbitrarily small $\phi_0 / M$ and $k / m_\phi$ (depicted in the inset panels).
        Results are scaled by $\phi_0 / M$ to facilitate interpreting results in FLRW spacetime (see
        the main text for details).
        Note that the horizontal axis corresponds to the \textit{physical} wave number
        $k / a$ in FLRW spacetime.
        Gray lines depict the trajectory of the physical horizon
        $H(t) = m_\phi / [a(t) / a_i]^2$ for three different initial
        amplitudes $\phi_{0,i} = M$, $\sqrt{5} M$, and $5 M$, roughly indicating the low--wave-number cutoff for resonance in the radiation era as a function of time.
    }
    \label{fig:Floq}
\end{figure*}

For massless dark photons, unstable solutions are present only for sizeable oscillation amplitudes,
$\phi_0\gtrsim M$.
While massive dark photons also experience these large-amplitude instabilities (provided the mass is not too large,
i.e., $\mA \lesssim \dot{\bar{W}} / \bar{W} \sim m_\phi$), the presence of the mass term  gives
rise to a novel low-momentum ($k\ll m_\phi$) instability in the small-amplitude ($\phi_0\ll M$)
regime when $\mA/m_\phi = 1/2$.
To see how this instability arises, we first note that in this limit and to leading order in
$\phi_0/M$, \cref{eqn:A0-eom-rescaled-minkowski,eqn:Apm-eom-rescaled-minkowski} tend to the Mathieu
equation
\begin{align}
    0
    &= \frac{\mathrm{d}^2 X}{\mathrm{d} z^2}
        + \left[ p - 2 q \cos(2z) \right] X(z),
    \label{eqn:mathieu}
\end{align}
where
\begin{subequations}
\label{eq:MathPar}
\begin{align}
    z
    &= \frac{m_\phi t}{2}
    \\
    p
    &= \left(\frac{2 k}{m_\phi}\right)^2
        + \left(\frac{2\mA}{m_\phi}\right)^2
    \\
    q
    &= \frac{2\phi_0}{M}\frac{\mA^2}{m_\phi^2}
        \times
        \begin{cases}
            1 & X= \mathcal{A}_0
            \\
            1-m_\phi^2/(2\mA^2) & X = \mathcal{A}_\pm.
        \end{cases}
    \label{eqn:mathieu-q}
\end{align}
\end{subequations}
The solution to the Mathieu equation~\cite{magnus2004hill},
$\tilde{\mathcal{P}}_+(t)e^{\tilde{\mu} z}+\tilde{\mathcal{P}}_-(z)e^{-\tilde{\mu} z}$, is unstable
for small $q$ provided that $p = n^2$ for integer $n$.
For instance, the case $n=1$, corresponding to $\mA=m_\phi/2$ for modes $k \ll m_\phi$, gives rise
to an unstable solution in the limit of small $q$ (and therefore of small amplitude).
The Floquet exponent of the Mathieu equation in this regime is known analytically to be
$\lim_{q\rightarrow0}\Re(\tilde{\mu}^{n=1})\approx \vert q \vert /2$~\cite{magnus2004hill},
yielding\footnote{
    Note that for $n=1$, the Mathieu resonance parameters $q$ [\cref{eqn:mathieu-q}]
    for $\mathcal{A}_\pm$ and $\mathcal{A}_0$ have opposite signs but the same magnitude
    (and therefore the same Floquet exponent).
}
\begin{align}
    \label{eq:FloqMink}
    \Re(\mu)\approx \frac{m_\phi}{8} \frac{\phi_0}{M}
\end{align}
for the solutions to \cref{eqn:A0-eom-rescaled-minkowski,eqn:Apm-eom-rescaled-minkowski} when $\phi_0\ll M$ and $k\ll m_\phi$.

\Cref{fig:Floq} depicts the Floquet exponents for the transverse
[\cref{eqn:Apm-eom-rescaled-minkowski}] and longitudinal [\cref{eqn:A0-eom-rescaled-minkowski}]
components as a function of the vector's wave number and the dilaton's oscillation amplitude.
In addition to the broad instability bands for large momenta and amplitudes, \cref{fig:Floq}
exhibits the above-described band at small $k / m_\phi$ extending down to $\phi_0/M \rightarrow 0$.
We note that the unstable band does not extend to arbitrarily small amplitudes if $\mA$ is not
precisely equal to $m_\phi/2$.
In particular, the width of the instability band in $k$ in the small-$q$
limit is $\vert p_{n=1} - 1 \vert = \vert q \vert$~\cite{magnus2004hill}.
Quantifying the mass tuning as
\begin{align}
    \delta
    &\equiv \left( \frac{\mA}{m_\phi} \right)^2 - \frac{1}{4},
\end{align}
the width of the instability band is
\begin{align}
    \left\vert
        \left( \frac{k}{m_\phi} \right)^2
        + \delta
    \right\vert
    &\lesssim \frac{\phi_0}{8 M}.
    \label{eqn:width-of-unstable-band}
\end{align}
The low-momentum instability band vanishes for $\phi_0 / M < 8 \delta$ (though if
$\delta < 0$, a narrow instability band persists to arbitrarily small $\phi_0 / M$ over wave numbers
satisfying $\vert \delta \vert - \phi_0 / 8 M \lesssim (k / m_\phi)^2 \lesssim \vert \delta \vert + \phi_0 / 8 M$).
Below we estimate the level of mass tuning this implies for viable dark photon dark matter
scenarios.

\subsection{Parametric resonance in FLRW}\label{sec:FLRW}

The above results for the resonant instabilities in Minkowski spacetime can be extended to the
expanding Universe by accounting for the redshifting of physical momenta $k / a$ and of the
dilaton's oscillation amplitude.
Specifically, in an FLRW spacetime, the dilaton begins to oscillate when $H(t_i) \approx m_\phi$ with an
initial amplitude $\phi_{0,i}$ that subsequently redshifts as\footnote{
    More precisely, the solution to \cref{eqn:dilaton-eom} for the homogeneous mode $\bar{\phi}(t)$
    in a radiation background is
    $\Gamma(5/4) \phi_{0, i} J_{1/4}(m_\phi t) / \sqrt[4]{m_\phi t / 2}$,
    where $J_\nu$ is the order-$\nu$ Bessel function.
    The asymptotic oscillation amplitude is
    $2^{3/2} \Gamma(5/4) \phi_{0, i} / (a / a_i)^{3/2} \sqrt{\pi}$.
}
\begin{align}
\label{eq:FLRWampl}
    \phi_0(t)
    &\approx 1.5 \phi_{0,i} \left( \frac{a(t)}{a_i} \right)^{-3/2}
\end{align}
in the radiation-dominated era.
The corresponding time dependence of the coefficients in the equations of motion
\cref{eqn:A0-eom-rescaled-minkowski,eqn:Apm-eom-rescaled-minkowski}, which nominally brings them
away from the form of the Mathieu equation \cref{eqn:mathieu}, is negligible when the dilaton's
oscillation rate is much faster than the expansion rate (which, in the radiation-dominated era,
evolves as $H \propto 1/a^2$).
[Likewise, the terms proportional to the Hubble rate in
\cref{eqn:A0-eom-rescaled,eqn:Apm-eom-rescaled} are also negligible in this limit.]
That is, at sufficiently late times, the dilaton's oscillation amplitude and each mode's physical
wave number are effectively constant over each oscillation.
The Minkowski-space Floquet exponent \cref{eq:FloqMink} therefore captures the instantaneous growth
rate in FLRW spacetime to good approximation upon replacing the constant $\phi_0$ with the (slowly)
decaying amplitude \cref{eq:FLRWampl}.

Depleting the dilaton's energy (so that the majority of the dark matter comprises dark photons)
requires parametric resonance to be efficient~\cite{Amin:2014eta}, i.e., that the exponential growth
rate is significantly greater than the Hubble rate ($\Re(\mu) / H \gg 1$) for a sufficiently long
duration.
Crucially, the expansion rate decays faster with expansion than the dilaton's oscillation amplitude
[\cref{eq:FLRWampl}] in the radiation-dominated era---namely, the exponential growth rate
\cref{eq:FloqMink} relative to the expansion rate is
\begin{align}
\label{eq:FlorFLRW}
    \frac{\Re[\mu(t)]}{H}
    &= \frac{1}{8} \frac{3 \phi_{0,i}}{2 M} \left( \frac{a(t)}{a_i} \right)^{1/2}.
\end{align}
Hence, the efficiency (i.e., per $e$-fold of expansion) of resonance grows with the expansion of the
Universe.
Accordingly, $\Re[\mu(t)] / H$ is approximately equal to the quantity plotted in \cref{fig:Floq}
times $\sqrt{a(t) / a_i} \cdot 3 \phi_{0, i} / 2 M$.
Furthermore, the comoving width of the instability band also grows with time, bounded
above by $(k / m_\phi)^2 \leq \phi_{0, i} / 8 M \cdot (a / a_i)^{1/2}$  [via
\cref{eqn:width-of-unstable-band}]
and below by the comoving horizon scale $m_\phi / (a / a_i)$ (corresponding to the gray lines in
\cref{fig:Floq}).

\Cref{eq:FlorFLRW} shows that, even if the initial oscillation amplitude is small, after the
Universe expands by a factor $\propto (\phi_{0,i} / M)^{-2}$, the rate of particle production becomes
efficient.
More precisely, production of the vector completes soon after its energy density becomes comparable
to the dilaton's.
Via \cref{eq:FLRWampl}, $\rho_\phi \approx m_\phi^2 \phi_{0,i}^2 / (a / a_i)^{3}$.
Since the vector is produced well after $a_i$ with a comoving wave number of order $m_\phi$, it is
nonrelativistic at production; its energy density is then approximately
\begin{align}\label{eqn:rho-A}
    \rho_A(t)
    &\approx \frac{\mA^2}{2 (a / a_i)^3 \bar{W}}
        \sum_{\lambda} \left\langle \mathcal{A}_\lambda(t, \mathbf{x})^2 \right\rangle,
\end{align}
where
\begin{align}
    \left\langle \mathcal{A}_\lambda(t, \mathbf{x})^2 \right\rangle
    &=  \int \mathrm{d} \ln k \,
        \frac{k^3}{2 \pi^2}
        \left\langle \vert \mathcal{A}_\lambda(t_i, k) \vert^2 \right\rangle
        e^{ 2 \int_{t_i}^{t} \mathrm{d} t' \, \Re[\mu(t)] }.
\end{align}
Before production, $\mathcal{A}_\lambda$ is in the vacuum state~\cite{Bunch:1978yq,Birrell:1982ix},
\begin{align}
    \left\langle
        \vert \mathcal{A}_\lambda(t_i, k) \vert^2
    \right\rangle
    &= \frac{1}{2 \sqrt{(k / a)^2 + \mA^2}}.
\end{align}
Setting the growth rate to that from \cref{eq:FlorFLRW} in \cref{eqn:rho-A},
taking the integral over wave number to be dominated by modes with $k \sim m_\phi$,
and noting that $\bar{W} \approx 1$ at late times,
\begin{align}
    \rho_A(t)
    &\approx \frac{\mA^2 m_\phi^2}{(a / a_i)^3}
        \exp\left[ \frac{3 \phi_{0, i}}{4 M} \left( \frac{a}{a_i} \right)^{1/2} \right].
\end{align}
The dilaton and vector have comparable energy density at a scale factor
\begin{align}
    \frac{a_\star}{a_i}
    &\equiv \left( \frac{3 \phi_{0, i}}{4 M} \right)^{-2}
        \ln \left(
            \frac{m_\phi^2}{\mA^2}
            \frac{\phi_{0, i}^2}{m_\phi^2}
        \right)^2.
    \label{eqn:astar}
\end{align}
For $\phi_{0, i} / M = 1$ and $\mA = 10^{-18} \, \mathrm{eV}$ (choosing $\phi_{0, i}$ to match the
relic abundance of dark matter---see \cref{sec:DMAbun}),
$a_\star / a_i \approx 7.1 \times 10^{4}$, decreasing only by a factor of $2.3$ for
$\mA = 10^{-6} \, \mathrm{eV}$.

The above results apply only while modes remain in the small-amplitude resonance band.
While the instability extends to arbitrarily small amplitudes when the vector's mass is precisely
half the dilaton's, resonance will terminate early if the mass ratio is not so precisely tuned.
Concretely, the $n = 1$ band must have nonzero width in $k$ at sufficiently small
$\vert q \vert \propto \phi_0 / M$.
Requiring that low-momentum modes do not leave the instability region by the time resonance
becomes efficient---namely, that \cref{eqn:width-of-unstable-band} remains satisfied for some range
of $k$ until the Universe expands by $a_\star / a$---we arrive at
\begin{align}
    \vert \delta \vert
    &< \frac{3 \phi_{0, i}}{16 M (a_\star / a_i)^{3/2}}
    \lesssim 10^{-7} \left( \frac{\phi_{0, i}}{2 M} \right)^{4}
\end{align}
for masses $\mA \sim 10^{-18} \, \mathrm{eV}$.

While other narrow-band resonances exist (e.g., if $\mA$ is slightly smaller than $2 m_\phi$ or for
other mass ratios $\mA = n m_\phi / 2$), they are centered at nonzero $k$ (in contrast to the
low-momentum $n = 1$ band) and shrink in width as the dilaton's amplitude decays.
The redshifting of momenta in FLRW spacetime therefore prevents any mode from growing large enough
to deplete the dilaton background of its energy density in such cases.

\section{Dark matter abundance}\label{sec:DMAbun}

We now determine the parameter space for which dark photons make up all of the dark matter.
The dark matter must have been produced sufficiently long before scales observable in the CMB
become dynamical, so production takes place deep in the radiation-dominated era.
The dilaton begins oscillating when $H \approx m_\phi$; at this stage, it contains a negligible
fraction of the total energy of the Universe (else the matter-dominated era begins too early).
Once the dilaton transfers a substantial fraction of its energy to the dark photons, the subsequent
evolution is nonlinear---in particular, the excited vector field modes rapidly backreact on the
dilaton background.
We assume for simplicity that most of the dilaton energy is transferred into the dark photons\footnote{
    Classical lattice simulations demonstrate efficient depletion of an oscillating dilaton into
    massless dark photons~\cite{Giblin:2017wlo,Deskins:2013dwa,Adshead:2017xll,Adshead:2018doq}
    in the strong coupling regime.
    Furthermore, in similar models with an axial coupling rather than a dilatonic one, most of the energy is transferred to the dark photons, regardless of whether they are massless~\cite{Adshead:2015pva,Adshead:2016iae,Figueroa:2017qmv,Cuissa:2018oiw,Adshead:2019igv,Adshead:2019lbr,Kitajima:2020rpm,Ratzinger:2020oct,Weiner:2020sxn} or not~\cite{Agrawal:2018vin}.
    Determining whether the same conclusion holds in the small-amplitude regime would
    require dedicated numerical simulations.
} so that the dilaton makes a negligible contribution to the dark matter abundance.
If the dark photons are nonrelativistic when produced,\footnote{
    As discussed previously, in the small-amplitude regime with $\mA = m_\phi / 2$, the dark photons are necessarily
    nonrelativistic at production, since $a_\star / a_i \gg 1$ via \cref{eqn:astar}.
    In the broad resonance regime (for any $\mA \lesssim m_\phi$ and sufficiently large
    $\phi_{0, i} / M > 1$), the dark photons are instead mildly relativistic at production, leading to
    a corresponding dilution in their abundance since they redshift more rapidly than matter
    until they become nonrelativistic.
    In this case, the predictions should be analogous to dark photons resonantly produced by
    axions~\cite{Agrawal:2018vin}.
} their abundance today is simply
\begin{align}\label{eqn:DPDMabundance}
    \frac{\Omega_{\gamma'} h^2}{0.12}
    &\approx
        \left( \frac{\mA}{10^{-17} \, \mathrm{eV}} \right)^{1/2}
        \left( \frac{\phi_{0, i}}{10^{16} \, \mathrm{GeV}} \right)^2,
\end{align}
taking $g_\star = 10.75$ effective number of relativistic degrees of freedom in the plasma at the
time of production.

CMB observations place two further constraints on parameter space.
First, ensuring that the dark matter was produced before scales observed in the CMB reenter the
horizon imposes a lower limit on the vector mass in the $\mA = m_\phi / 2$, small-amplitude regime
because the Universe expands by a substantial amount before resonance becomes efficient.
Namely, using conservation of entropy and \cref{eqn:astar}, the redshift of production
$z_\star = a_0 / a_\star - 1$ satisfies
\begin{align}
    \frac{z_\star + 1}{1.9 \times 10^{5}}
    &= \left( \frac{\phi_{0, i}}{M}\right)^{2}
        \left(
            \frac{\mA}{10^{-17} \, \mathrm{eV}}
        \right)^{1/2}
\end{align}
(dropping additional logarithmic dependence on $m_\phi$ and $\phi_{0, i}$).
Since the redshift of matter-radiation equality is $\approx 3400$~\cite{Planck:2018vyg}, for masses
$\mA \lesssim 10^{-17} \, \mathrm{eV}$, dark photons are not produced early enough
if $\phi_{0, i} / M < 1$.
However, lighter masses can be accommodated by a modest increase in amplitude
$\phi_{0, i} / M \propto \mA^{-1/4}$.

Second, in this scenario, the vacuum expectation value of the dilaton arises from quantum
fluctuations during inflation.
The value of $\phi_{0,i}$ varies across causally disconnected Hubble patches and generates an
isocurvature perturbation after the decay into dark photons.
The power of the isocurvature perturbation from inflation is
$\mathcal{P}_S=H_I^2/(\pi\phi_{0,i})^2$, where $H_I$ is the Hubble scale during inflation.
Assuming $\Omega_{\gamma'}=\Omega_\mathrm{DM}$, CMB constraints on isocurvature
perturbations~\cite{Planck:2018jri} bound the Hubble scale during inflation to be below
\begin{align}
    H_I
    < 3 \times 10^{11} \, \mathrm{GeV}
        \left( \frac{\mA}{10^{-17} \mathrm{eV}} \right)^{-1/4}.
\end{align}

\section{Discussion and Conclusions}

\label{sec:Concl}

We have demonstrated that a massive vector field kinetically coupled to an oscillating scalar field
exhibits a novel nonlinear decay channel.
Namely, in addition to a strong-coupling regime reminiscent of axion--dark-photon
models~\cite{Agrawal:2018vin,Co:2018lka}, an oscillating dilaton twice as heavy as its coupled dark
photon induces substantial vector production even for small oscillation amplitudes.
This effect allows for the efficient production of dark photon dark matter without invoking large
couplings, naturally giving rise to dark photon dark matter over a wide range of masses.
Efficient parametric resonance in the small-coupling regime requires a finely tuned mass ratio
$\mA/m_\phi=1/2$---as severe as one part in $10^7$ for order-unity couplings---barring a UV model
that gives rise to both the dark photon and dilaton's masses and explains the coincidence.

In \cref{sec:DMAbun} we showed that requiring consistency with the CMB jointly constrains the vector
mass and coupling strength, but only modestly large couplings are required to achieve ultralight
masses $\sim 10^{-20}$ to $10^{-18} \, \mathrm{eV}$.
To go beyond these estimates (e.g., to characterize the relic abundance of dark photons and dilatons
as a function of the particle masses and initial field amplitudes and to compute the resulting
spectra of dark photons) requires $3+1$D numerical simulations.
Preliminary simulations confirm the numerical and parametric estimates of
\cref{sec:FLRW,sec:DMAbun}, but we leave a more thorough investigation of these questions to future
work.

The rapid growth of vector modes and nonlinear dynamics at the end of the production phase would
source a gravitational wave background, providing a possible probe of the model.
Though a quantitative prediction would again require dedicated numerical study, the signals are
likely to resemble those from dark photons resonantly produced by a rolling
axion~\cite{Machado:2018nqk, Machado:2019xuc, Weiner:2020sxn, Ratzinger:2020oct, Salehian:2020dsf}.
However, the signal amplitude is unlikely to be promisingly large.
Stochastic backgrounds are parametrically suppressed by both the fraction of Universe's net energy
density contained by the source and the gravitational-wave wavelength relative to the horizon size
at production~\cite{Giblin:2014gra}.
In this scenario, the vector and dilaton are massive relics whose abundance is
$\propto a_\star / a_\mathrm{eq}$, i.e., the ratio of the scale factors at production and
matter-radiation equality.
One might then expect the long delay from oscillation ($a_i$) to production ($a_\star$) to enhance
the resulting signals, but gravitational wave emission occurs at fixed comoving scales $\sim m_\phi$
that are $a_\star / a_i$ times farther inside the horizon at production.
These two effects turn out to cancel each other, and even highly efficient gravitational wave
production is unlikely to exceed present-day abundances of $\Omega_{\mathrm{GW}, 0} \sim 10^{-15}$
(see, e.g., Ref.~\cite{Cyncynates:2022wlq}).

Here we specifically invoke a dilatonic coupling to achieve dark matter predominantly comprising
dark photons.
In principle, the dilaton could instead be the dark matter and simply happen to have such a coupling
with parameter values that fail to achieve efficient conversion to dark photons before the CMB
forms.
For the same reason that resonance could become efficient at sufficiently late times in the
radiation era even for small couplings [as discussed after \cref{eq:FlorFLRW}], in this alternative
scenario, substantial dark photon production will \textit{never} occur: As the Universe transitions
to matter domination, the Hubble rate instead decays as $a^{-3/2}$ just like the dilaton, and the
growth rate per Hubble time asymptotes to a constant.
In other words, if the dilaton has not efficiently produced dark photons by the time the CMB forms,
then it never will.
(The marginal regime---resonance becoming efficient as CMB modes enter the horizon---would likely be incompatible with the cold, collisionless dark matter required by CMB observations.)\footnote{
    We thank an anonymous referee for pointing out this interesting feature of the model.
}

While we have not assumed any particular origin for the mass of the dark photon, in the case where
the mass arises from the Higgs mechanism, the production of vortices challenges the viability of
dark photon dark matter~\cite{East:2022rsi}.
Moreover, as noted by~\cite{Goodsell:2009xc, Reece:2018zvv, Agrawal:2018vin}, Stueckelberg masses
are restricted to $\mA \gtrsim$ meV in string theory, implying that smaller masses must arise
from the Higgs mechanism.
To our knowledge, all proposed production mechanisms for light dark photon dark matter are afflicted
by vortex formation constraints.
However, it is possible that the small-amplitude regime we point out could evade such issues: The
long delay between dilaton oscillations and dark photon production could ensure the vector's energy
density never exceeds the critical value for vortex formation.
We defer a full exploration of the parameter dependence of vortex formation constraints, along with
the phenomenology of kinetic mixing with the Standard Model photon and subsequent plasma effects, to
future work.\footnote{
    A kinetic mixing term would nominally induce resonant production of Standard Model photons
    (depending on, e.g., its plasma mass) as well, but the kinetic mixing parameter must be (much)
    smaller than $10^{-6}$ to $10^{-10}$ (depending on the dark photon mass)~\cite{Caputo:2021eaa},
    greatly suppressing this effect.
}

Throughout this work, we assumed that the dark photon mass is $\phi$ independent. This dependence,
like the form of the kinetic coupling $W(\phi)$, is ultimately determined by the UV physics.
Providing a UV completion of the theory is beyond the scope of this work, but it is conceivable that
both the kinetic and mass terms attain a $\phi$ dependence, coming from, e.g., radiative
corrections.
[In fact, the small-amplitude resonance is present if the coupling function $W(\phi)$ multiplies the
mass term rather than the kinetic term of the vector since the pertinent terms in the equation of
motion have the same form.]
Such interactions, as well as dark photon self-interactions, could have consequences for the
efficiency of dark photon production and the viable parameter space for dark photon dark matter.
We leave the investigation of such effects for future work.

\begin{acknowledgments}
We thank Mustafa Amin and David Cyncynates for comments on a draft of this paper,
and Z.J.W.\ also thanks David Cyncynates for extensive discussions about dark photons.
P.A.\ thanks the Center for Particle Cosmology at the University of Pennsylvania for hospitality while this work was being completed.
The work of P.A.\ and K.L.\ is supported in part by the United States Department of Energy, DE-SC0015655.
Z.J.W.\ is supported by the Department of Physics and the College of Arts and Sciences at the University of Washington.
This work made use of the Python packages \textsf{NumPy}~\cite{Harris:2020xlr},
\textsf{SciPy}~\cite{Virtanen:2019joe}, \textsf{matplotlib}~\cite{Hunter:2007ouj},
\textsf{SymPy}~\cite{Meurer:2017yhf}, and
\textsf{CMasher}~\cite{cmasher}.
\end{acknowledgments}

\bibliography{references}

\end{document}